\documentclass[twocolumn,showpacs,floats,floatfix,superscriptaddress,aps,pra]{revtex4-1}
\usepackage{amsfonts}
\usepackage{amssymb}
\usepackage{amsmath}
\usepackage{calc}
\usepackage{graphicx}
\usepackage{bm}

\usepackage[normalem]{ulem}
\usepackage{amsmath,amssymb}
\usepackage{multirow}
\usepackage{xcolor,soul}
\usepackage{srcltx}
\usepackage{hyperref,graphicx}

\def\be{ \begin{equation}}
\def\ee{ \end{equation}}
\def\bea{ \begin{eqnarray}}
\def\eea{ \end{eqnarray}}
\def\bse{ \begin{subequations}}
\def\ese{ \end{subequations}}
\def\bc{ \begin{center}}
\def\ec{ \end{center}}

\def\H{\mathbf{H}}

\def\c{\mathbf{c}}
\def\a{\mathbf{a}}

\def\R{\mathbf{R}}

\def\sech{\,\text{sech}}
\def\ket#1{\vert #1 \rangle}

\def\half{\tfrac12}

\begin{document}

\author{Boyan T. Torosov}
\altaffiliation{Permanent address: Institute of Solid State Physics, Bulgarian Academy of Sciences, 72 Tsarigradsko chauss\'{e}e, 1784 Sofia, Bulgaria}
\affiliation{Dipartimento di Fisica, Politecnico di Milano and Istituto di Fotonica e Nanotecnologie del Consiglio Nazionale delle Ricerche, Piazza L. da Vinci 32, I-20133 Milano, Italy}
\author{Giuseppe Della Valle}
\affiliation{Dipartimento di Fisica, Politecnico di Milano and Istituto di Fotonica e Nanotecnologie del Consiglio Nazionale delle Ricerche, Piazza L. da Vinci 32, I-20133 Milano, Italy}
\author{Stefano Longhi}
\affiliation{Dipartimento di Fisica, Politecnico di Milano and Istituto di Fotonica e Nanotecnologie del Consiglio Nazionale delle Ricerche, Piazza L. da Vinci 32, I-20133 Milano, Italy}
\title{Non-Hermitian shortcut to stimulated Raman adiabatic passage}
\date{\today}

\begin{abstract}
We propose a non-Hermitian  generalization of stimulated Raman adiabatic passage (STIRAP), which allows one to increase speed and fidelity of the adiabatic passage. This is done by adding balanced imaginary (gain/loss) terms in the diagonal (bare energy) terms of the Hamiltonian and choosing them such that they cancel exactly the nonadiabatic couplings, providing in this way  an effective shortcut to adiabaticity. Remarkably, for a STIRAP using delayed Gaussian-shaped  pulses in the counter-intuitive scheme the imaginary terms of the Hamiltonian turn out to be time independent. A possible physical realization of non-Hermitian STIRAP, based on light transfer in three evanescently-coupled optical waveguides, is proposed.
\end{abstract}

\maketitle

\section{Introduction}

Stimulated Raman adiabatic passage (STIRAP) is one of the most popular tools, used for manipulation of quantum structures \cite{STIRAP}. This method transfers population adiabatically between two states $\ket{1}$ and $\ket{3}$, in a three-level quantum system, without populating the intermediate state $\ket{2}$. The applications of STIRAP cover a huge part of contemporary physics: coherent atomic excitation \cite{Bruce}, control of chemical reactions \cite{chemical}, quantum information processing \cite{STIRAP-QIP}, coherent quantum state transfer and spatial adiabatic passage \cite{uffa}, waveguide optics \cite{waveguides}, to name a few. 

The technique of STIRAP is based on the existence of a dark state, which is an eigenstate of the Hamiltonian and is a time-dependent superposition of the initial and target states. Because STIRAP is an adiabatic technique, it achieves high fidelity only in the limit of adiabatic evolution, which requires large temporal pulse areas. If the adiabatic condition is not perfectly realized, the nonadiabatic coupling between the eigenstates causes the efficiency of the process to drop down.
Several methods to achieve higher fidelity of STIRAP have been proposed. One scenario is to use an additional field to cancel the nonadiabatic coupling \cite{short1}. Another approach is to minimize the nonadiabatic losses by applying Lewis-Riesenfeld inverse engineering \cite{inverse}, which has been shown in \cite{inverse2} to be potentially equivalent to Berry's transitionless quantum driving \cite{Berry}. One may also use the Dykhne-Davis-Pechukas formula which leads to the so called parallel adiabatic passage \cite{PAP}. These techniques imply a strict time dependence of the pulse shapes and detuning. Finally, one can use composite pulses to improve dramatically the fidelity of adiabatic passage \cite{CP}, by choosing appropriately the relative phases of the pulses in the sequence. Other theoretical techniques proposed to engineer the shortcuts, with a detailed discussion of the experimental results and prospects, can be found in the recent review \cite{reviewMuga}.

In the recent years there has been a growing interest in the use of non-Hermitian (NH) Hamiltonians \cite{Moiseyev}, especially in the context of $\mathcal{PT}$-symmetric systems \cite{PT,PT2}.  It was demonstrated, for instance, that a $\mathcal{PT}$-symmetric Hamiltonian can produce a faster than Hermitian evolution in a two-state quantum system, while keeping the eigenenergy difference fixed \cite{Faster}. NH extensions have been done on the classical Landau-Zener model \cite{LZNH} and some schemes for the realisation of $\mathcal{PT}$-symmetry have been proposed \cite{PT-exper}. Recently, an approximation of the adiabatic condition for NH systems was also derived \cite{AdiabaticNH}. 

In this paper, we propose a technique to increase the speed of STIRAP by using a NH term in the Hamiltonian to cancel the nonadiabatic coupling; hence we call this method a NH shortcut to STIRAP. The proposed method is similar to the NH shortcut for two-level adiabatic processes, studied in \cite{NH-Torosov}.
The paper is organized as follows. First, we shortly review the theory of STIRAP in the case of resonant excitation. Then we study how additional NH terms in the Hamiltonian affect the evolution and we show how NH terms can be used to cancel the nonadiabatic coupling. We then consider two concrete examples and examine the limitations of our method. Interestingly, for a STIRAP using delayed Gaussian-shaped  pulses in the counter-intuitive scheme the imaginary energy terms of the Hamiltonian turn out to be time independent.
Finally, a possible physical implementation of NH shortcut of STIRAP in waveguide optics is briefly discussed.


\section{Theory of STIRAP in an Hermitian system}

\begin{figure}[tb]
\includegraphics[width=8.5cm]{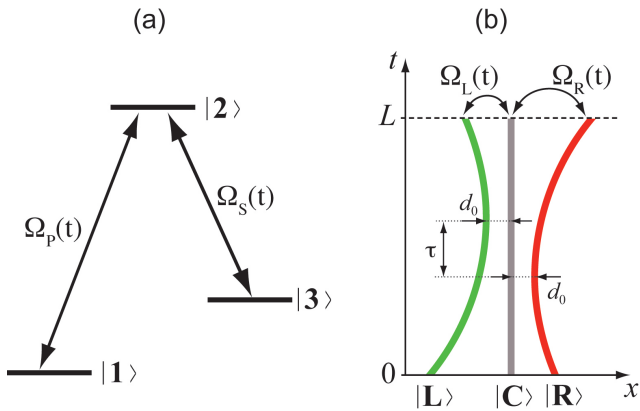}
\caption{(a) STIRAP coupling scheme. The three states $\ket{1}$, $\ket{2}$ and $\ket{3}$ form a lambda configuration. The pump field $\Omega_p(t)$ couples states $\ket{1}$ and $\ket{2}$ and the Stokes field couples states $\ket{3}$ and $\ket{2}$. (b) Tunneling-coupled optical waveguides realizing NH STIRAP in the counter-intuitive scheme for Gaussian pulses. Populations of atomic states $\ket{1}$, $\ket{2}$ and  $\ket{3}$ in (a) are mimicked by light power propagating in waveguides $\ket{L}$, $\ket{C}$ and $\ket{R}$, respectively. Waveguide $\ket{L}$ is lossy with loss rate $\gamma$, whereas waveguide $\ket{R}$ provides gain $-\gamma$. The pump (Stokes) Rabi frequency $\Omega_{p(s)}(t)$ corresponds to the tunneling rate $\Omega_{L(R)}(t)$ of outer waveguides with the center waveguide.}
\label{system}
\end{figure}

We start by briefly reviewing the theory of STIRAP \cite{STIRAP}. Let us consider a three-level quantum system in a lambda configuration [Fig.~\ref{system}(a)], coherently driven by two external fields, the pump field $\Omega_p(t)$, and the Stokes field $\Omega_s(t)$. The goal is to transfer all the population from the initial state $\ket{1}$ to the final state $\ket{3}$, without populating the intermediate state $\ket{2}$. The evolution of the system is described by the Schr\"{o}dinger equation 
\be\label{Schr}
i \hbar\partial_t \mathbf{c}(t) = \H(t)\mathbf{c}(t),
\ee
where the vector $\c(t) = [c_1(t), c_2(t), c_3(t)]^T$ contains the three probability amplitudes of states $\ket{1}$, $\ket{2}$ and $\ket{3}$.
The Hamiltonian in the rotating-wave approximation and assuming exact resonance reads \cite{STIRAP}
\be\label{H}
\H(t) = \frac\hbar 2 \left[ \begin{array}{ccc} 0  & \Omega_{p}(t) & 0 \\
\Omega_{p}(t)  & 0  & \Omega_{s}(t)\\
0 & \Omega_{s}(t)& 0
\end{array} \right],
\ee
where the Rabi frequencies $\Omega_p(t)$ and $\Omega_s(t)$ of the pump and Stokes pulses are assumed to be real. The mechanism of the population transfer in STIRAP and the limitation of adiabaticity can be easily understood if we introduce the so-called adiabatic basis. This is the basis, formed of the instantaneous eigenstates 
$\ket{\Phi_{+}(t)}$, $\ket{\Phi_{0}(t)}$, $\ket{\Phi_{-}(t)}$
of the time-varying Hamiltonian $\H(t)$. These \emph{adiabatic states} are connected to the original states $\ket{1}$, $\ket{2}$ and $\ket{3}$ by 
\bse\label{adBasis}
\begin{align}
&\ket{\Phi_{+}(t)}= \frac{1}{\sqrt{2}}\left(\sin\theta\ket{1}+\ket{2}+\cos\theta\ket{3}\right),\\ \label{adState}
&\ket{\Phi_{0}(t)}= \cos\theta\ket{1}-\sin\theta\ket{3}, \\
&\ket{\Phi_{-}(t)}= \frac{1}{\sqrt{2}}\left(\sin\theta\ket{1}-\ket{2}+\cos\theta\ket{3}\right),
\end{align}
\ese
where the time-dependent mixing angle $\theta(t)$ is defined as 
\be
\tan\theta(t) = \Omega_p(t)/\Omega_s(t).
\ee
The probability amplitudes of the adiabatic states $\a(t)=[a_+(t), a_0(t), a_-(t)]^T$ are connected to the original ones by using the transformation
\be
\c(t)=\R(t)\a(t) ,
\ee 
where the transformation matrix $\R$ is given by
\be\label{R}
\R(t) = \frac{1}{\sqrt{2}} \left[ \begin{array}{ccc} \sin\theta  & \sqrt{2}\cos\theta & \sin\theta \\
1  & 0  & -1\\
\cos\theta & -\sqrt{2}\sin\theta & \cos\theta
\end{array} \right] .
\ee
The Schr\"{o}dinger equation in the adiabatic basis can be written as
\be\label{SchrA}
i \hbar\partial_t \mathbf{a}(t) = \H_a(t)\mathbf{a}(t),
\ee
with $\H_a=\R^{-1}\H\R -i\hbar\R^{-1}\dot{\R}$, or explicitly
\be\label{Ha}
\H_a(t) = \frac{\hbar}{\sqrt{2}} \left[ \begin{array}{ccc} \frac{\Omega}{\sqrt{2}}  & i\dot{\theta} & 0 \\
-i\dot{\theta}  & 0  & -i\dot{\theta}\\
0 & i\dot{\theta}& -\frac{\Omega}{\sqrt{2}}
\end{array} \right],
\ee
where $\Omega(t)=\sqrt{\Omega^2_p(t)+\Omega^2_s(t)}$. Now, if we assume an adiabatic (slow) evolution, then $\dot{\theta} \simeq 0$ which means that if the system is initially in one of the adiabatic states, then it stays in it during the evolution. It is this type of adiabatic following, which is in the essence of STIRAP. Let us now take a closer look at the adiabatic state $\ket{\Phi_0(t)}$. We see that it contains only states $\ket{1}$ and $\ket{3}$ and has no component on state $\ket{2}$ [see Eq.~\eqref{adState}]. Furthermore, for a \emph{counterintuitive} order of the pump and Stokes pulses, we have the relations $\Omega_p/\Omega_s \stackrel{t\to -\infty}{\longrightarrow} 0$ and $\Omega_p/\Omega_s \stackrel{t\to \infty}{\longrightarrow} \infty$. Hence, as time evolves, the mixing angle $\theta$ rises from $0$ to $\pi/2$ and we get 
\be
\ket{\Phi_0(-\infty)}=\ket{1},\quad \ket{\Phi_0(\infty)}=-\ket{3}.
\ee
This means, that if the evolution is perfectly adiabatic, and the system is initially prepared in state $\ket{1}$, at the end of the process we will achieve a complete population transfer to state $\ket{3}$, with negligible excitation of the intermediate state $|2 \rangle$. However, the evolution is never perfectly adiabatic and some nonadiabatic coupling $i\dot{\theta}$ is always present, which limits the efficiency of STIRAP. In the next section we show how this problem can be overcome by adding NH terms in the Hamiltonian.

\section{Non-Hermitian shortcut}

We consider again the STIRAP Hamiltonian \eqref{H}, but this time we add two time-dependent NH terms $\pm i\gamma$ in the diagonal \cite{note},

\be\label{Hg}
\H^\gamma(t) = \frac\hbar 2 \left[ \begin{array}{ccc} -i\gamma(t)  & \Omega_{p}(t) & 0 \\
\Omega_{p}(t)  & 0  & \Omega_{s}(t)\\
0 & \Omega_{s}(t)& i\gamma(t)
\end{array} \right].
\ee
which correspond to time-varying complex energies of the bare states $|1\rangle$ and $|3 \rangle$.
In the adiabatic basis \eqref{adBasis}, this Hamiltonian has the form
\begin{align}
\label{Hga}\notag
&\H^\gamma_a(t) = \\
&\hbar \left[ \begin{array}{ccc} 
\frac{\Omega}{2}+i\frac{\gamma \cos2\theta}{4}   & -i\frac{\gamma\sin2\theta}{2\sqrt{2}}+i\frac{\dot{\theta}}{\sqrt{2}} & i\frac{\gamma \cos2\theta}{4}  \\
 -i\frac{\gamma\sin2\theta}{2\sqrt{2}}-i\frac{\dot{\theta}}{\sqrt{2}}  & -i\frac{\gamma \cos2\theta}{2}  & -i\frac{\gamma\sin2\theta}{2\sqrt{2}}-i\frac{\dot{\theta}}{\sqrt{2}}\\
i\frac{\gamma \cos2\theta}{4}  &  -i\frac{\gamma\sin2\theta}{2\sqrt{2}}+i\frac{\dot{\theta}}{\sqrt{2}} & -\frac{\Omega}{2}+i\frac{\gamma \cos2\theta}{4}
\end{array} \right].
\end{align}
Our goal is to choose $\gamma(t)$ in  such a way that the terms in the Hamiltonian \eqref{Hga}, which cause the evolution of the system to deviate from state $\ket{\Phi_0(t)}$, vanish. This can be achieved by setting
\be\label{gamma}
\gamma(t)=\frac{2\dot{\theta}(t)}{\sin 2\theta(t)},
\ee
which nullifies the first and the third element from the second column of $\H^\gamma_a$. In this way the evolution of the amplitudes of the adiabatic state obey the equations
\bse
\begin{align}
&a_+(t_f)=0 ,\\ 
&a_0(t_f)=\exp\left[ -\half\int_{t_i}^{t_f}\gamma(t)\cos 2\theta(t)dt \right], \label{A0}\\
&a_-(t_f)=0 ,
\end{align}
\ese
where we have assumed that initially the state of the system coincides with the adiabatic state $\ket{\Phi_0}$, hence $a_0(t_i)=1$.
It can be shown that if $\Omega_p(t)$ and $\Omega_s(t)$ are reflections of each other, $\Omega_p(t)=\Omega_s(\tau - t)$ [e.g., if $\Omega_p(t)$ and $\Omega_s(t)$ are identical symmetric functions of time], where $\tau$ is the pulse delay, then $\gamma(t)$ is an even function of time and $\cos2\theta(t)$ is an odd function of time. Hence, if we assume symmetric time interval, $t_f=-t_i$, the integral in \eqref{A0} is equal to zero. This means that despite the fact that the Hamiltonian is NH, the norm of the state vector at the end of the evolution will be equal to unity. There is, however, an essential distinction between the Hermitian and non-Hermitian method in the transient population during the evolution. While in the Hermitian case the norm of the state vector remains unity during the entire evolution, in the NH STIRAP, as seen from Eq.~\eqref{A0}, the norm of the state vector will change in time.

\section{Examples}

\begin{figure}[tb]
\includegraphics[width=15cm]{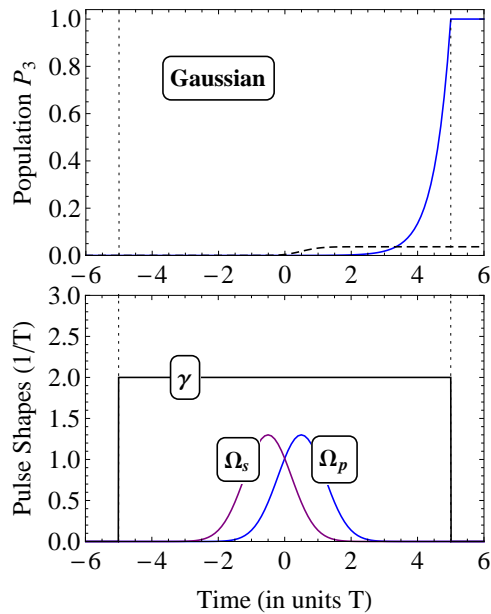}
\caption{(Top) Transition probability $P_{1\to 3}$ as a function of time for Gaussian pulse shapes \eqref{gaussians} with $\Omega_0=1.3/T$ and $\tau=T$. The black dashed line illustrates the standard STIRAP technique, the solid blue line shows the NH STIRAP, and the vertical dotted lines mark the initial and final moments of time $t_i$ and $t_f$, used in the integration of the Schr\"{o}dinger equation. (Bottom) Shapes of the pump and Stokes pulses and of the gain/loss function $\gamma$(t).}
\label{populationGaus}
\end{figure}

In order to better illustrate the method of the NH shortcut of STIRAP, we consider two special cases, and we calculate the NH (gain/loss) functions $\gamma(t)$ for both of them. As a first example, let us assume that the pump and Stokes pulses have Gaussian shape,
\bse\label{gaussians}
\begin{align}
&\Omega_p(t)=\Omega_0 e^{-(t-\tau/2)^2/T^2},\\
&\Omega_s(t)=\Omega_0 e^{-(t+\tau/2)^2/T^2},
\end{align}
\ese
where $\Omega_0$ is the peak Rabi frequency, $T$ is the pulse duration and $\tau$ is the delay between the pulses.
For this pulse shape, it can be easily shown that the gain/loss function of Eq.~\eqref{gamma} is constant,
\be\label{eq15}
\gamma(t)=2\frac{\tau}{T^2} .
\ee
This time independence of $\gamma$ allows one a relatively easy implementation of the NH Hamiltonian in waveguide optics, where $\gamma$ describes optical amplification/attenuation in the outer waveguides of a three-waveguide structure \cite{waveguides}, as discussed in more detail in the next section.

In Fig.~\ref{populationGaus} we plot the evolution in time of the population of the target state $\ket{3}$, for the regular STIRAP ($\gamma=0$) and for the NH extension ($\gamma\neq 0$). We can see from the figure, that the NH STIRAP strongly outperforms the original Hermitian STIRAP, by achieving complete population transfer, while for the standard STIRAP we have only $P_3 \approx 0.05$ for the same values of $\Omega_0$ and $\tau$.  

\begin{figure}[tb]
\includegraphics[width=15cm]{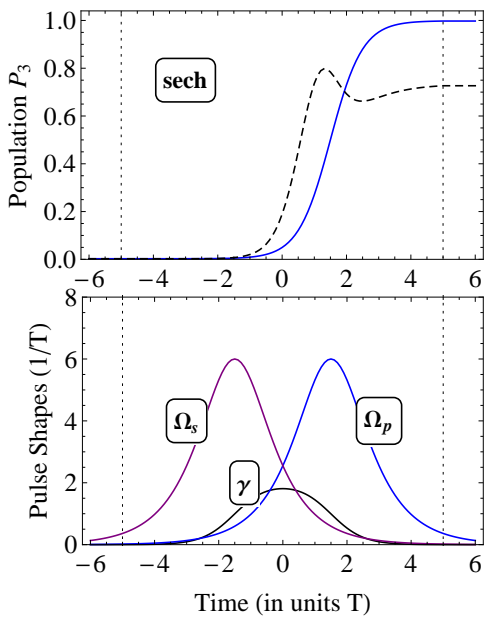}
\caption{Same as Fig.~\ref{populationGaus} but for hyperbolic secant pulse shapes \eqref{sech} with $\Omega_0=6/T$ and $\tau=3T$.}
\label{populationSech}
\end{figure}

As a second example we assume hyperbolic secant pulse shapes,
\bse\label{sech}
\begin{align}
&\Omega_p(t)=\Omega_0 \sech \left( \frac{t-\tau/2}{T} \right),\\
&\Omega_s(t)=\Omega_0 \sech \left( \frac{t+\tau/2}{T} \right).
\end{align}
\ese
In this case, the gamma term is given by
\be
\gamma(t)=\frac{1}{T}\left[\tanh\left(\frac{t+\tau/2}{T}\right)-\tanh\left(\frac{t-\tau/2}{T}\right)\right].
\ee
In Fig.~\ref{populationSech} we show the evolution of the population of the target state, as well as the pulse shapes and $\gamma$. Again, we find that by using the NH STIRAP, we obtain much higher fidelity, even for parameter values where the ordinary STIRAP fails. 

We want to note here, that both in the cases of Gaussian and sech pulse shapes, the function $\gamma(t)$ does not depend on the particular value of the peak Rabi frequency. This is an interesting feature, which allows our technique to be used even for small values of $\Omega_0$, where the usual STIRAP technique would not work.

\section{Feasible implementation in optics}

\begin{figure}[tb]
\includegraphics[width=8.8cm]{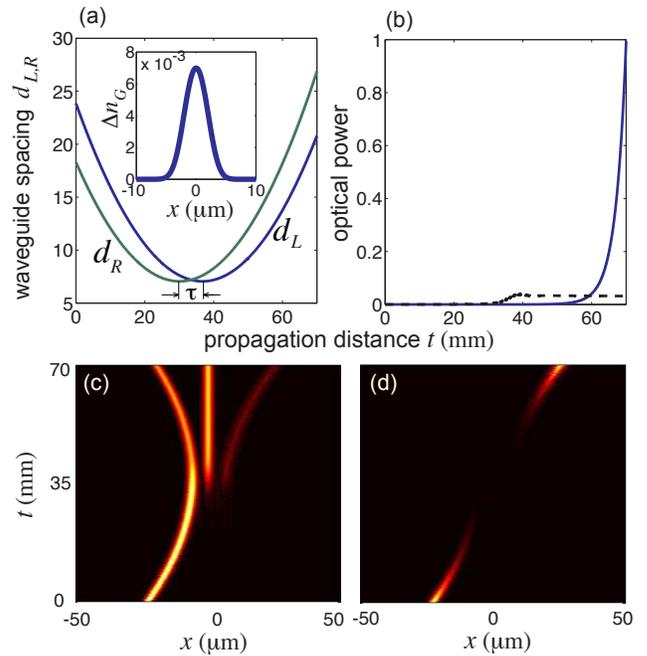}
\caption{ Beam propagation in the three-waveguide optical system of Fig.~\ref{system}(b) that realizes the NH Hamiltonian \eqref{Hg}. (a) Behavior of the waveguide separation distances $d_L$ and $d_R$ of waveguides $|L \rangle$ and $|R \rangle$ from the central waveguide $|C \rangle$. The refractive index profile $\Delta n_G(x)$ of the single waveguide channel used in the numerical simulations is depicted in the inset. (b) Evolution of the optical beam power trapped in waveguide $|R \rangle$, normalized the optical power that excites waveguide $|L \rangle$ at the input plane. The dashed curve refers to the Hermitian model (i.e. absence of loss and gain in waveguides $|L \rangle$ and $|R \rangle$), whereas the solid curve corresponds to the NH model with a uniform loss/gain rate defined by Eq.~\eqref{eq15}. Panels (c) and (d) show  the evolution of beam intensity along the waveguide system, as obtained by solving the optical Schr\"{o}dinger equation \eqref{Schr-opt} using a beam propagation method, in the Hermitian [panel (c)] and non-Hermitian [panel (d)] case. Parameter values are given in the text.}
\label{fig4}
\end{figure}

As anticipated above, our suggested method of NH shortcut to adiabaticity could be physically realized in waveguide optics \cite{waveguides}. In this case, the system is composed of three optical waveguides $\ket{L}$, $\ket{C}$ and $\ket{R}$ in the geometrical setting schematically shown in Fig.~\ref{system}(b). In photonic STIRAP, the role of Rabi frequencies $\Omega_p $ and $\Omega_s $  is played by the tunneling rates $\Omega_L$ and $\Omega_R$ between the center waveguide $\ket{C}$ and the waveguides $\ket{L}$ and $\ket{R}$ respectively. Therefore, to implement the desired pulse shape, the separation distance $d_L$ and $d_R$ between outer and center waveguides is slowly varied along the propagation direction $t$.
Since the tunneling rate scales exponentially with the waveguide separation according to the law $\Omega_{L,R}(t) =( \Omega_0/2) \exp\{-k[d_{L,R}(t)-d_0]\}$ (with $\Omega_0$ and $k$ two constant parameters to be determined from the fabrication process), in order to implement a Gaussian pulse shape a quadratic law for $d_L$ and $d_R$ is required (see Ref.~\citenum{waveguides} for further details). In our NH generalization of optical STIRAP, for a Gaussian pulse shape, a longitudinally-independent optical loss $-\gamma$ and an equal rate of net optical amplification $\gamma$ should be applied to the outer waveguides $\ket{L}$ and $\ket{R}$, respectively. A possible implementation of such an active/passive system can be achieved in Ti in-diffused Fe:LiNbO$_3$ optical waveguides, which have been recently exploited to experimentally demonstrate $\mathcal{PT}$-symmetric directional couplers in the visible \cite{Ruter}. Actually, the geometry of Fig.~\ref{system}(b) can be easily obtained in Ti in-diffused LiNbO$_3$ waveguides by using standard photolithography to shape the Ti stripes before in-diffusion. Moreover, loss rate $\gamma$ can be controlled by the amount of Fe$^{2+}$ dopants, giving rise to optical
excitation of electrons from Fe$^{2+}$ centers to the conduction band of LiNbO$_3$. Finally, optical gain is provided under laser pumping through two-wave mixing process thanks to the photorefractive nonlinearity induced by Fe-doping: a gain coefficient $g$ of few cm$^{-1}$ can be easily achieved ($g = 3.8$~cm$^{-1}$ in Ref.~\citenum{Ruter}). Note that, as in Ref.~\citenum{Ruter}, the pump laser beam is incident from the top of the sample and a suitable amplitude mask ought to be employed to selectively control the amount of pump illumination in the three waveguides so to achieve: $g = 0$, i.e. no pumping, to have net attenuation with rate $-\gamma$ in $\ket{L}$; $g = \gamma$, i.e. transparency, in $\ket{C}$; and $g = 2\gamma$, so to have net optical amplification $g-\gamma=\gamma$, in $\ket{R}$. To provide more evidence of feasibility of our scheme, we numerically studied beam propagation in a three-waveguide optical structure in the geometrical setting schematically shown in Fig.~\ref{system}(b). In the scalar and paraxial approximations, the optical Schr\"{o}dinger equation for the electric field envelope $\psi(x,t)$ that describes beam propagation in the optical structure reads (see, for instance, \cite{LonghiLPR})
\begin{equation}\label{Schr-opt}
i \lambdabar \frac{ \partial \psi}{\partial t}=- \frac{\lambda^2}{2n_s} \frac{\partial^2 \psi}{\partial x^2}+V(x,t) \psi
\end{equation}
where $\lambdabar= \lambda/ (2 \pi)$, $\lambda$ is the wavelength of light wave propagating in the dielectric medium, $n_s$ is the substrate refractive index at wavelength $\lambda$, $t$ and $x$ are the longitudinal and transverse spatial coordinates, respectively, and $V(x,t)$ is the optical potential, which is related to the refractive index change $\Delta n(x,t)$ of the guiding structure from the substrate region by the simple relation $V(x,t)=-\Delta n(x,t)$. An error function shape $\Delta n_G(x)=\Delta n_0 \{ {\rm erf}[(x+w_1)/D_x]-{\rm erf} [(x-w_1)/D_x)] \} / [2 \; {\rm erf} (w_1/D_x)]$  has been assumed for the refractive index profile of each of the three waveguides  in the structure [see the inset of Fig.~\ref{fig4}(a)], with a channel width $2w_1=4 \; \mu$m, a diffusion length $D_x=2 \; \mu$m and a maximum index change $\Delta n_0=7 \times 10^{-3}$, which are typical to Ti in-diffused LiNbO$_3$ optical waveguides. To simulate loss and gain in the outer waveguides $|L \rangle$ and $|R \rangle$, a small imaginary part $\Delta n_I$  of opposite signs, leading to a complex refractive index,  has been added to $\Delta n_0$ for the waveguides  $|L \rangle$ and $|R \rangle$.  The optical structure is probed at $\lambda=514$ nm (Ar-ion laser light), corresponding to a bulk refractive index $n_s=2.33$. The outer waveguides $|L \rangle$ and $|R \rangle$ are parabolically-curved along the propagation direction $t$;  the detailed behavior of the spacings $d_L$ and $d_R$ of waveguides $|L \rangle$ and $|R \rangle$ from the straight central waveguide $|C \rangle$ are depicted in  Fig.~\ref{fig4}(a). The minimum separation distance is $d_0=5.95 \; \mu$m, whereas the longitudinal shift is $\tau=7 \;$mm [see Fig.~\ref{system}(a)]. The bent profiles of waveguides  $|L \rangle$ and $|R \rangle$ yield an effective Gaussian variation with $t$ of the coupling constants $\Omega_L$ and $\Omega_R$ [see Eqs.~\eqref{gaussians}], corresponding to $\Omega_0 \simeq 1.985 \; {\rm cm}^{-1}$ and $T \simeq \tau= 7 \;$mm, i.e. $\Omega_0 T=1.3$ as in Fig.~\ref{populationGaus}(a). The full sample length is $L=10T=70 \; $mm.
Figure \ref{fig4}(c) shows, in a pseudo-color map, the numerically-computed light intensity evolution along the sample  for $\Delta n_I=0$, i.e. in the absence of gain and loss in the outer waveguides.  At the input plane, $t=0$,  the waveguide $|L \rangle$ is excited by a Gaussian beam which is well overlapped with the fundamental mode of the waveguide.
The corresponding behavior of the optical light power trapped in waveguide $|R \rangle$, normalized to the input optical power value in waveguide $|L \rangle$, is shown by the dashed curve of Fig.~\ref{fig4}(b), which reproduces very well the behavior predicted by coupled-mode equations (the dashed curve in the upper panel of Fig.~\ref{populationGaus}). To cancel the non-adiabatic terms, according to Eq.~\eqref{eq15} a power loss/gain coefficient $\gamma=2 \tau / T^2 \simeq 2.8 \; {\rm cm}^{-1}$ is  required. Such a gain/loss value, which can be realized by Fe-doping, is obtained by assuming an imaginary part $\Delta n_I \simeq \pm 0.0016 \Delta n_0$ of the refractive index change in waveguides  $|L \rangle$ and  $|R \rangle$. In Fig.~\ref{fig4}(d) we show  the numerically-computed light intensity evolution along the sample in the presence of loss and gain in waveguides  $|L \rangle$ and  $|R \rangle$, respectively. The corresponding behavior of the normalized optical light power trapped in waveguide $|R \rangle$ is shown by the solid curve in Fig.~\ref{fig4}(b), which reproduces very well the curve predicted by the coupled-mode equation model (the solid curve in the upper panel of Fig.\ref{populationGaus}).

\section{Conclusions and discussion}

In this paper we have studied theoretically a NH generalization of STIRAP, which allows one to cancel the nonadiabatic coupling and increase the speed and fidelity of the process. We have examined two special cases, which allow to understand how the technique works for a particular shape of the pump and Stokes pulses. Remarkably, for a STIRAP using delayed Gaussian-shaped  pulses in the counter-intuitive scheme, the required imaginary terms of the Hamiltonian turn out to be time independent. This allowed us to propose a feasible implementation of NH STIRAP in optical waveguides with uniform gain/loss.

We want to note here, that our method cannot be applied for arbitrary pulse shapes, but has some restrictions.
In order to find the domain of applicability of our technique, we can express the gain/loss function $\gamma$ in terms of the pump and Stokes Rabi frequencies.
Using Eq.~\eqref{gamma}, after some simple algebraic calculations, we obtain
\be
\gamma(t)=\frac{\dot{\Omega}_p(t)}{\Omega_p(t)}-\frac{\dot{\Omega}_s(t)}{\Omega_s(t)} =\frac{d}{dt}\ln\frac{\Omega_p(t)}{\Omega_s(t)}
\ee
Hence, in order to be able to apply this method, the logarithm of the ratio of the two pulse functions needs to have well defined first derivative. For instance, if we try to use $\sin^2$ pulse shapes, one can easily check that the $\gamma$ term is divergent, which obviously limits the applicability of our method.

Another point, which is important to stress is that the technique of NH shortcut is very sensitive to the initial conditions. The initial state of the system must be very closely prepared to the adiabatic state $\ket{\Phi_0}$. Otherwise, the norm of the state vector will not be conserved and some extra gain or loss will be introduced.

Finally, we point out that the scheme, which we propose, only works for the case of resonant STIRAP. If the one-photon detuning is not equal to zero, the nonadiabatic couplings will depend on two mixing angles; hence it is not straightforward to nullify the losses and further investigation may be necessary. One possible direction for further investigation, in order to overcome this issue, might be to use the non-Hermitian generalization of the Lewis-Resenfield theory \cite{NH-Lewis} to find a shortcut for such Hamiltonian.

\acknowledgments

This work was supported by the Fondazione Cariplo (Grant No. 2011-0338).



\begin{thebibliography}{99}

\bibitem{STIRAP} 
K. Bergmann, H. Theuer, and B. W. Shore, Rev. Mod. Phys. \textbf{70}, 1003 (1998);
N. V. Vitanov, M. Fleischhauer, B. W. Shore, and K. Bergmann, Adv. At., Mol., Opt. Phys. \textbf{46}, 55 (2001);
N. V. Vitanov, T. Halfmann, B. W. Shore, and K. Bergmann, Annu. Rev. Phys. Chem. \textbf{52}, 763 (2001).

\bibitem{Bruce} B. W. Shore, \textit{The Theory of Coherent Atomic Excitation} (Wiley, New York, 1990).

\bibitem{chemical} P. Dittmann, F. P. Pesl, J. Martin, G. W. Coulston, G. Z. He and K. Bergmann, J. Chem. Phys. \textbf{97}, 9472 (1992);

\bibitem{STIRAP-QIP}
 M. Hennrich, T. Legero, A. Kuhn, and G. Rempe, Phys. Rev. Lett. \textbf{85}, 4872 (2000);
 A. Kuhn, M. Hennrich, and G. Rempe, ibid. \textbf{89}, 067901 (2002);
 T. Wilk, S. C. Webster, H. P. Specht, G. Rempe, and A. Kuhn, ibid. \textbf{98}, 063601 (2007);
 J. L. S{\o}rensen, D. M{\o}ller, T. Iversen, J.B. Thomsen, F. Jensen, P. Staanum, D. Voigt, and M. Drewsen, New J. Phys. \textbf{8}, 261 (2006).
 
\bibitem{uffa} 
K. Eckert, M. Lewenstein, R. Corbalan, G. Birkl, W. Ertmer, and
J. Mompart, Phys. Rev. A {\bf 70}, 023606 (2004); A. D. Greentree, J. H. Cole, A. R. Hamilton, and L. C. L. Hollenberg,
Phys. Rev. B {\bf 70}, 235317 (2004); L. C. L. Hollenberg, A. D. Greentree, A. G. Fowler, and C. J.
Wellard, Phys. Rev. B {\bf 74}, 045311 (2006); K. Eckert, J. Mompart, R. Corbalan, M. Lewenstein, and G. Birkl,
Opt. Commun. {\bf 264}, 264  (2006).

\bibitem{waveguides} E. Paspalakis, Opt. Commun. \textbf{258}, 30 (2006);
S. Longhi, Phys. Rev. E \textbf{73}, 026607 (2006);
S. Longhi, G. Della Valle, M. Ornigotti, and P. Laporta, Phys. Rev. B \textbf{76}, 201101 (2007);
Y. Lahini, F. Pozzi, M. Sorel, R. Morandotti, D. N. Christodoulides, and Y. Silberberg, Phys. Rev. Lett. \textbf{101}, 193901 (2008); 
S.-Y. Tseng and M.-C. Wu, IEEE Photon. Technol. Lett. {\bf 22}, 1211 (2010);
T.-Y. Lin, F.-C. Hsiao, Y.-W. Jhang, C. Hu, and
S.-Y. Tseng, Opt. Express {\bf 20}, 24085 (2012);  
A. P. Hope, T. G. Nguyen, A. D. Greentree, and A. Mitchell, Opt. Express {\bf 21}, 22705 (2013).


\bibitem{short1} R. G. Unanyan, L. P. Yatsenko, B. W. Shore, K. Bergmann, Opt. Commun. \textbf{139}, 48 (1997); 
Xi Chen, I. Lizuain, A. Ruschhaupt, D. Gu{\'e}ry-Odelin, and J. G. Muga, Phys. Rev. Lett. \textbf{105}, 123003 (2010);
L. Giannelli and E. Arimondo, Phys. Rev. A \textbf{89}, 033419 (2014).
 
\bibitem{inverse} Xi Chen and J. G. Muga, Phys. Rev. A \textbf{86}, 033405 (2012).

\bibitem{inverse2} Xi Chen, E. Torrontegui, and J. G. Muga, Phys. Rev. A \textbf{83}, 062116 (2011).

\bibitem{Berry} M. V. Berry, J. Phys. A \textbf{42}, 365303 (2009).
 
\bibitem{PAP} G. Dridi, S. Gu{\'e}rin, V. Hakobyan, H. R. Jauslin, and H. Eleuch, Phys. Rev. A \textbf{80}, 043408 (2009).

\bibitem{CP} B. T. Torosov, S. Gu\'{e}rin, and N. V. Vitanov, Phys. Rev. Lett. \textbf{106}, 233001 (2011).
B. T. Torosov and N. V. Vitanov, Phys. Rev. A \textbf{87}, 043418 (2013).

\bibitem{reviewMuga}
E. Torrontegui, S. Ib\'{a}\~{n}ez, S. Martinez-Garaot, M. Modugno, A. del Campo, D. Guery-Odelin, A. Ruschhaupt, Xi Chen, and J. G. Muga, Adv. At. Mol. Opt. Phys. {\bf 62}, 117 (2013).

\bibitem{Moiseyev}
N. Moiseyev, {\it Non-Hermitian Quantum Mechanics} (Cambridge University Press, London,  Cambridge, 2011).

\bibitem{PT} C. M. Bender and S. Boettcher, Phys. Rev. Lett. \textbf{80}, 5243 (1998);
C. M. Bender, Rep. Prog. Phys. {\bf 70}, 957 (2007).

\bibitem{PT2}
A. Mostafazadeh, J. Math.Phys. {\bf 43}, 205 (2002); A. Mostafazadeh, J. Phys. A {\bf 36}, 7081 (2003).

\bibitem{Faster} C. M. Bender, D. C. Brody, H. F. Jones, and B. K. Meister, Phys. Rev. Lett. \textbf{98}, 040403 (2007);
R. Uzdin, U. G\"{u}nther, S. Rahav, and N. Moiseyev, J. Phys. A: Math. Theor. \textbf{45}, 415304 (2012).

\bibitem{LZNH} E. M. Graefe, H. J. Korsch, Czech. J. Phys. \textbf{56}, 1007 (2006); 
S. A. Reyes, F. A. Olivares and L. Morales-Molina, J. Phys. A: Math. Theor. \textbf{45}, 444027 (2012); 
R. Uzdin and N. Moiseyev, J. Phys. A: Math. Theor. \textbf{45}, 444033 (2012).

\bibitem{PT-exper} C. Hang, G. Huang, and V. V. Konotop, Phys. Rev. Lett. \textbf{110}, 083604 (2013);
H. Li, J. Dou, and G. Huang, Opt. Express 21, \textbf{32053} (2013).

\bibitem{AdiabaticNH} S. Ib\'{a}\~{n}ez and J. G. Muga, Phys. Rev A \textbf{89}, 033403 (2014). 

\bibitem{NH-Torosov} B. T. Torosov, G. Della Valle, and S. Longhi, Phys. Rev. A \textbf{87}, 052502 (2013).

\bibitem{note}
Note that, at each time instant $t$, the NH matrix $\mathbf{H}^{\gamma}(t)$ turns out to be $\mathcal{PT}$-symmetric if $\Omega_1(t)=\Omega_2(t)$ [see: 
C. M. Bender, P. N. Meisinger, and Q. Wang, J. Phys. A {\bf 36}, 6791 (2003)]. To realize NH STIRAP, a counter-intuitive sequence of Rabi frequencies is needed, so that in our case the NH matrix   $\mathbf{H}^{\gamma}(t)$  in not $\mathcal{PT}$ invariant.

\bibitem{Ruter}
C. E. R\"{u}ter, K. G. Makris, R. El-Ganainy, D. N. Christodoulides, M. Segev, and D. Kip, Nature Phys. {\bf 6}, 192 (2010).

\bibitem{LonghiLPR}
S. Longhi, Laser \& Photon. Rev. {\bf 3}, 243
(2009); I.L. Garanovich, S. Longhi, A.A. Sukhorukova, and Y.S. Kivshar, Phys. Rep. {\bf 518}, 1 (2012).

\bibitem{NH-Lewis} Xiao-Chun Gao, Jing-Bo Xu and Tie-Zheng Qian, Phys. Rev. A \textbf{46}, 3626 (1992).

\end{thebibliography}
\end{document}